\documentclass[conference, a4paper]{IEEEtran}
\IEEEoverridecommandlockouts
\usepackage{cite}
\usepackage{amsmath,amssymb,amsfonts}
\usepackage{algorithmic}
\usepackage{graphicx}
\usepackage{textcomp}
\usepackage{xcolor}
\def\BibTeX{{\rm B\kern-.05em{\sc i\kern-.025em b}\kern-.08em
    T\kern-.1667em\lower.7ex\hbox{E}\kern-.125emX}}

\newtheorem{theorem}{\it Theorem}
\newtheorem{lemma}{\it Lemma}

\newtheorem{corollary}{\it Corollary}
\newtheorem{proposition}{\it Proposition}

\begin{document}
	

\title{Generic Variance Bounds on Estimation and Prediction Errors in Time Series Analysis:\\An Entropy Perspective\\
\thanks{The work is partially supported by the Knut and Alice Wallenberg Foundation, the Swedish Strategic Research Foundation, the Swedish Research Council, the JSPS under Grant-in-Aid for Scientific Research Grant
	No. 18H01460, the NSF under grants CNS-1544782 and ECCS-1847056, and the ARO under grant W911NF1910041.}
}


\author{Song Fang, Mikael Skoglund, Karl Henrik Johansson, Hideaki Ishii, and Quanyan Zhu
	\thanks{Song Fang, Mikael Skoglund, and Karl Henrik Johansson are with the School of Electrical Engineering and Computer Science, KTH Royal Institute of Technology, Sweden, {\tt\small sonf@kth.se, skoglund@kth.se, kallej@kth.se}}%
	\thanks{Hideaki Ishii is with the Department of Computer Science, Tokyo Institute of Technology, Japan, {\tt\small ishii@c.titech.ac.jp}}%
	\thanks{Quanyan Zhu is with the Department of Electrical and Computer Engineering, New York University, USA, {\tt\small quanyan.zhu@nyu.edu}}%
}


\maketitle

\begin{abstract}
	
	In this paper, we obtain generic bounds on the variances of estimation and prediction errors in time series analysis via an information-theoretic approach. It is seen in general that the error bounds are determined by the conditional entropy of the data point to be estimated or predicted given the side information or past observations. Additionally, we discover that in order to achieve the prediction error bounds asymptotically, the necessary and sufficient condition is that the ``innovation" is asymptotically white Gaussian. When restricted to Gaussian processes and $1$-step prediction, our bounds are shown to reduce to the Kolmogorov--Szeg\"o formula and Wiener--Masani formula known from linear prediction theory.  

\end{abstract}

%

\section{Introduction}


The theory of linear prediction has been an important aspect of time series analysis, and it has broad applications in control/estimation, from Wiener filtering to Kalman filtering, and signal processing \cite{wiener1949extrapolation, bode1950simplified, kalman1960new, kailath1974view, makhoul1975linear, anderson2012optimal, pourahmadi2001foundations, vaidyanathan2007theory, caines2018linear}. The Kolmogorov--Szeg\"o formula \cite{pourahmadi2001foundations, Pap:02, vaidyanathan2007theory} and Wiener--Masani formula \cite{wiener1957prediction, lindquist2015linear, chen2018role} are two fundamental results that provide lower bounds on the variances of prediction errors for the linear prediction of Gaussian processes. In addition, the connections of these two formulae with entropy and information theory have been earlier established and utilized in, e.g., \cite{vaidyanathan2007theory, kim2010feedback} (see also the references therein).

In this paper, we go beyond the linear Gaussian case; we employ information theory as the main mathematical tool to obtain generic bounds on the variances of estimation and prediction errors in time series analysis, by investigating the underlying entropic relationships of the data points composing the time series. 
The fundamental error bounds are applicable to any causal estimators/predictors, while the processes to be estimated/predicted can have arbitrary distributions. More specifically, the error bounds are characterized explicitly by the conditional entropy of the data point to be estimated or predicted given the side information or past observations, that is, by the amount of ``randomness" contained in the data point when knowing the side information or past observations. As such, if the side information or past observations provide more/less information of the data point to be estimated or predicted, then the conditional entropy becomes smaller/larger, and thus the error bounds become smaller/larger.

Moreover, we show that in order to achieve the fundamental prediction error bounds asymptotically, the necessary and sufficient condition is that the ``innovation" \cite{linearestimation} is asymptotically white Gaussian. This mandates that the optimal, minimum-variance predictor should be an innovation ``Gaussianizing-whitening" predictor. In a broad sense, the innovation being white Gaussian means there is no ``information" remaining in the residue; in other words, all the information that may be utilized to reduce the variance of the error has been extracted in the Gaussianizing-whitening procedure. This observation coincides with the conclusions in Wiener filtering and Kalman filtering \cite{linearestimation, FangACC18}, where the innovations associated with the optimal filters are white Gaussian.


In the special case of Gaussian processes and $1$-step (ahead) prediction, our bounds reduce to the well-known Kolmogorov--Szeg\"o formula and Wiener--Masani formula. However, to the best of our knowledge, such formulae for $m$-step ($m > 1$) prediction and/or non-Gaussian processes have been lacking. Our corresponding bounds derived indicate that there will be an additional term of mutual information in the case of $m$-step prediction compared with $1$-step prediction, and this additional term never decreases (and in most cases, increases) as $m$ becomes larger. On the other hand, the bounds obtained for $1$-step prediction and non-Gaussian processes manifest that there will be an additional term as a function of the so-called negentropy rate in the case of non-Gaussian processes compared with Gaussian processes, and this additional term decreases as the processes to be predicted become less Gaussian.

The remainder of the paper is organized as follows. Section~II introduces the technical preliminaries. In Section~III, we introduce one by one the variance bounds on estimation errors, the variance bounds on prediction errors, their connections with the Kolmogorov--Szeg\"o formula as well as Wiener--Masani formula, and further discussions. Concluding remarks are given in Section~IV.

\section{Preliminaries}

Throughout this paper, we consider real-valued continuous random variables and vectors, as well as discrete-time stochastic processes. All the random variables, random vectors, and stochastic processes are assumed to be zero-mean. We represent random variables and vectors using boldface letters. Given a stochastic process $\left\{ \mathbf{x}_{k}\right\}$, we denote the sequence $\mathbf{x}_0,\ldots,\mathbf{x}_{k}$ by the random vector $\mathbf{x}_{0,\ldots,k}=\left[\mathbf{x}_0^T,\ldots,\mathbf{x}_{k}^T\right]^T$ for simplicity. The logarithm is defined with base $2$. All functions are assumed to be measurable. 

A stochastic process $\left\{ \mathbf{x}_{k}\right\}, \mathbf{x}_{k} \in \mathbb{R}^m$ is said to be asymptotically stationary if it is stationary as $k \to \infty$, and herein stationarity means strict stationarity unless otherwise specified \cite{Pap:02}. In addition, a process being asymptotically stationary implies that it is asymptotically mean stationary \cite{gray2011entropy}.
The asymptotic power spectrum of a zero-mean asymptotically stationary process $\left\{ \mathbf{x}_{k} \right\}$ is defined as
\begin{flalign}
&\Phi_{\mathbf{x}}\left( \omega\right)
=\sum_{k=-\infty}^{\infty} R_{\mathbf{x}}\left( k\right) \mathrm{e}^{-\mathrm{j}\omega k}, \nonumber 
\end{flalign}
where 
$R_{\mathbf{x}}\left( k\right) =\lim_{i\to \infty} \mathrm{E}\left[ \mathbf{x}_i \mathbf{x}^{T}_{i+k} \right]$ denotes the asymptotic correlation matrix. 
It is known that $\Phi_{\mathbf{x}}\left( \omega\right)$ is positive semidefinite. Moreover, the asymptotic covariance of $\left\{ \mathbf{x}_{k}\right\}$ can be computed as
\begin{flalign}
\lim_{k\to \infty} \mathrm{E}\left[ \mathbf{x}_k\mathbf{x}^{T}_{k} \right]
= R_{\mathbf{x}}\left( 0\right) 
= \frac{1}{2\pi}\int_{-\pi}^{\pi} \Phi_{\mathbf{x}}\left( \omega\right) \mathrm{d}  \omega. \nonumber
\end{flalign}

Definitions and properties of the information-theoretic notions that will be used in this paper, such as differential entropy $h\left( \mathbf{x} \right)$, conditional differential entropy $
h\left(\mathbf{x} \middle| \mathbf{y}\right)$, entropy rate $h_\infty \left(\mathbf{x}\right)$, and mutual information $I\left(\mathbf{x};\mathbf{y}\right)$, can be found in, e.g., \cite{Pin:64, Pap:02, Cov:06}.

The estimation counterpart to Fano's inequality is given below \cite{Cov:06}.

\begin{lemma}
	Given any random variable $\mathbf{x} \in \mathbb{R}$ with differential entropy $h \left( \mathbf{x} \right)$, and letting $\overline{\mathbf{x}}$ be an estimate of $\mathbf{x}$, the variance of the estimation error must satisfy
	\begin{flalign}
	\mathrm{E} \left[ \left( \mathbf{x} - \overline{\mathbf{x}} \right)^2 \right]  \geq \frac{1}{2 \pi \mathrm{e}} 2^{2 h \left( \mathbf{x} \right)}. \nonumber
	\end{flalign}
\end{lemma}

Note that herein and for the rest of this paper, we assume the estimate is unbiased without loss of generality, since the minimum-variance estimate is unbiased \cite{Cov:06}. This means that the error variance bounds obtained for unbiased estimators are also valid for biased estimators; stated alternatively, they are optimal among all (biased and unbiased) estimators. 

With side information, the following corollary holds \cite{Cov:06}.

\begin{lemma} \label{side}
	Given any random variable $\mathbf{x} \in \mathbb{R}$ and its side information $\mathbf{y}$, the variance of the estimation error for any estimator $\overline{\mathbf{x}} = f \left( \mathbf{y} \right)$ must satisfy
	\begin{flalign}
	\mathrm{E} \left[ \left( \mathbf{x} - \overline{\mathbf{x}} \right)^2 \right] \geq \frac{1}{2 \pi \mathrm{e}} 2^{2 h \left( \mathbf{x} | \mathbf{y} \right)}, \nonumber
	\end{flalign}
	where $h \left( \mathbf{x} | \mathbf{y} \right)$ denotes the conditional entropy of $\mathbf{x}$ given $\mathbf{y}$.
\end{lemma}


The Kolmogorov--Szeg\"o formula \cite{pourahmadi2001foundations, Pap:02, vaidyanathan2007theory} indicates the minimum variance of prediction error for stationary Gaussian processes in Wiener filtering \cite{wiener1949extrapolation}.

\begin{lemma} \label{KS}
	Consider a stationary Gaussian process $\left\{ \mathbf{x}_{k}\right\}, \mathbf{x}_{k} \in \mathbb{R}$ with power spectrum $S_{\mathbf{x}} \left( \omega \right)$. Let $\overline{\mathbf{x}}_k$ denote the $1$-step ahead prediction (in the rest of the paper, ``$1$-step ahead prediction" will be abbreviated as ``$1$-step prediction" for simplicity) of $\mathbf{x}_k$ given $\mathbf{x}_{0, \ldots, k-1}$, and let $ \mathbf{x}_k - \overline{\mathbf{x}}_k$ denote the prediction error. Then, the minimum asymptotic variance of the prediction error is given by
	\begin{flalign}
	\min \lim_{k\to \infty} \mathrm{E}\left[ \left( \mathbf{x}_k - \overline{\mathbf{x}}_k \right)^2 \right] 
	= 2^{\frac{1}{2 \mathrm{\pi}}\int_{-\mathrm{\pi}}^{\mathrm{\pi}}{\log S_{\mathbf{x}} \left( \omega \right)  \mathrm{d}\omega }}. \nonumber
	\end{flalign}
	Herein, the minimum is taken over all causal predictors.
\end{lemma}

In the multivariate case, the following Wiener--Masani formula \cite{wiener1957prediction, lindquist2015linear, chen2018role} holds.

\begin{lemma} \label{WM}
	Consider a stationary Gaussian process $\left\{ \mathbf{x}_{k}\right\}, \mathbf{x}_{k} \in \mathbb{R}^m$ with power spectrum $\Phi_{\mathbf{x}} \left( \omega \right)$. The minimum of the determinant of the  asymptotic prediction error covariance satisfies
	\begin{flalign}
	& \min \lim_{k\to \infty} \det \mathrm{E}\left[ \left( \mathbf{x}_k - \overline{\mathbf{x}}_k \right) \left( \mathbf{x}_k - \overline{\mathbf{x}}_k \right)^T \right] \nonumber \\
	&~~~~ = 2^{\frac{1}{2 \mathrm{\pi}}\int_{-\mathrm{\pi}}^{\mathrm{\pi}}{\log \det \Phi_{\mathbf{x}} \left( \omega \right)  \mathrm{d}\omega }}. \nonumber
	\end{flalign}
	Herein, the minimum is taken over all causal predictors.
\end{lemma}

\section{Variance Bounds on Estimation and Prediction Errors} \label{null}

In this section, we introduce in order variance bounds on estimation errors, variance bounds on prediction errors, and their connections with the Kolmogorov--Szeg\"o formula as well as Wiener--Masani formula. We also provide further discussions that relate our results to various lines of research.

\subsection{Variance Bounds on Estimation Errors} \label{estimation}

We first provide a generic estimation error bound for when estimating a stochastic process with side information.

\begin{theorem} \label{MIMOFano}
	Consider a stochastic process $\left\{ \mathbf{x}_{k} \right\}, \mathbf{x}_{k} \in \mathbb{R}^{n}$ with side information $\left\{ \mathbf{y}_{k} \right\}$. Denote the estimation of $\mathbf{x}_{k}$ by $\overline{\mathbf{x}}_{k} = f_{k} \left( \mathbf{y}_{0,\ldots,k} \right)$.
	Then, the following estimation error bound always holds:  
\begin{flalign} \label{MIMOestimation}
\det \mathrm{E} \left[ \left( \mathbf{x}_{k} - \overline{\mathbf{x}}_{k} \right) \left( \mathbf{x}_{k} - \overline{\mathbf{x}}_{k} \right)^T \right]
\geq \frac{1}{\left( 2 \pi \mathrm{e} \right)^n} 2^{2 h \left( \mathbf{x}_k | \mathbf{y}_{0,\ldots,k} \right)},
\end{flalign}
where equality holds if and only if $\mathbf{x}_{k} - \overline{\mathbf{x}}_{k}$ is Gaussian and $I \left( \mathbf{x}_k - \overline{\mathbf{x}}_{k}; \mathbf{y}_{0,\ldots,k} \right) = 0$.
\end{theorem}

The proof is omitted due to lack of space; hereinafter, refer to the extended version \cite{FangITW19arxiv} of this paper for the proofs omitted in this version.

It is seen herein that the lower bound is solely determined by the conditional entropy of the data point $\mathbf{x}_{k}$ to be estimated given the side information $\mathbf{y}_{0,\ldots,k}$, that is, by the amount of ``randomness" contained in the data point when knowing the side information. As such, if the side information provides more/less knowledge of the data point to be estimated, then the conditional entropy becomes smaller/larger, and thus the error bound becomes smaller/larger.

In a broad sense, the term $\mathbf{x}_{k} - \overline{\mathbf{x}}_{k}$ can be viewed as the ``innovation" \cite{linearestimation} associated with the estimate $ \overline{\mathbf{x}}_{k}$. Hence, equality in \eqref{MIMOestimation} holds if and only if the innovation is Gaussian and contains no information of the side information $\mathbf{y}_{0,\ldots,k}$, i.e., $I \left( \mathbf{x}_k - \overline{\mathbf{x}}_{k}; \mathbf{y}_{0,\ldots,k} \right) = 0$. Intuitively, it is as if all the ``information" that may be utilized to reduce the estimation error variance has been extracted.



In the scalar case, the next corollary follows as a special case of Theorem~\ref{MIMOFano}.

\begin{corollary} \label{process}
	Consider a stochastic process $\left\{ \mathbf{x}_{k} \right\}, \mathbf{x}_{k} \in \mathbb{R}$ with side information $\left\{ \mathbf{y}_{k} \right\}$. Denote the estimation of $\mathbf{x}_{k}$ by $\overline{\mathbf{x}}_{k} = f_{k} \left( \mathbf{y}_{0,\ldots,k} \right)$. Then, the following estimation error bound always holds:  
	\begin{flalign} 
	\mathrm{E} \left[ \left( \mathbf{x}_{k} - \overline{\mathbf{x}}_{k} \right)^2 \right]
	\geq \frac{1}{2 \pi \mathrm{e}} 2^{2 h \left( \mathbf{x}_k | \mathbf{y}_{0,\ldots,k} \right)},
	\end{flalign}
	where equality holds if and only if $\mathbf{x}_{k} - \overline{\mathbf{x}}_{k}$ is Gaussian and $I \left( \mathbf{x}_k - \overline{\mathbf{x}}_{k}; \mathbf{y}_{0,\ldots,k} \right) = 0$.
\end{corollary}

Corollary~\ref{process} generalizes the estimation counterpart to Fano's inequality, as presented in Lemma~\ref{side}, from random variables to stochastic processes. Note that when the side information is $\mathbf{y}_{k}$ and $\overline{\mathbf{x}}_{k} = f_{k} \left( \mathbf{y}_{k} \right)$, Corollary~\ref{process} reduces to Lemma~\ref{side}.

Similarly, as a special case of Theorem~\ref{MIMOFano}, it may be shown that given any random vector $\mathbf{x} \in \mathbb{R}^n$ and its side information $\mathbf{y}$, the covariance of the estimation error for any estimator $\overline{\mathbf{x}} = f \left( \mathbf{y} \right)$ must satisfy
	\begin{flalign}
	\det \mathrm{E} \left[ \left( \mathbf{x} - \overline{\mathbf{x}} \right) \left( \mathbf{x} - \overline{\mathbf{x}} \right)^T \right]
	\geq \frac{1}{\left( 2 \pi \mathrm{e} \right)^n } 2^{2 h \left( \mathbf{x} | \mathbf{y} \right)},
	\end{flalign}
	where equality holds if and only if $\mathbf{x} - \overline{\mathbf{x}}$ is Gaussian and $I \left( \mathbf{x} - \overline{\mathbf{x}}; \mathbf{y} \right) = 0$. In addition, without side information $\mathbf{y}$, it holds that 
	\begin{flalign}
	\det \mathrm{E} \left[ \left( \mathbf{x} - \overline{\mathbf{x}} \right) \left( \mathbf{x} - \overline{\mathbf{x}} \right)^T \right]
	\geq \frac{1}{\left( 2 \pi \mathrm{e} \right)^n } 2^{2 h \left( \mathbf{x} \right)}.
	\end{flalign} 

\subsection{Variance Bounds on Prediction Errors}

In what follows, we introduce a generic prediction error bound for when predicting a process based on its past observations. The proof is again omitted. 

\begin{theorem} \label{MIMOprediction}
	Consider a stochastic process $\left\{ \mathbf{x}_{k} \right\}, \mathbf{x}_{k} \in \mathbb{R}^{n}$. Denote the $1$-step prediction of $\mathbf{x}_{k}$ by $\overline{\mathbf{x}}_{k} = f_{k} \left( \mathbf{x}_{0,\ldots,k-1} \right)$.
	Then, the following prediction error bound always holds:  
	\begin{flalign} \label{MIMOprediction1}
	\det \mathrm{E} \left[ \left( \mathbf{x}_{k} - \overline{\mathbf{x}}_{k} \right) \left( \mathbf{x}_{k} - \overline{\mathbf{x}}_{k} \right)^T \right]
	\geq \frac{1}{\left( 2 \pi \mathrm{e} \right)^n} 2^{2 h \left( \mathbf{x}_k | \mathbf{x}_{0,\ldots,k-1} \right)},
	\end{flalign}
    where equality holds if and only if $\mathbf{x}_{k} - \overline{\mathbf{x}}_{k}$ is Gaussian and $I \left( \mathbf{x}_k - \overline{\mathbf{x}}_{k}; \mathbf{x}_{0,\ldots,k-1} \right) = 0$.
\end{theorem}

Note that the prediction error bound is merely determined by the conditional entropy of the data point $\mathbf{x}_{k}$ to be predicted given the past observations $\mathbf{x}_{0,\ldots,k-1}$. In addition, equality in \eqref{MIMOprediction1} holds if and only if the current innovation $\mathbf{x}_k - \overline{\mathbf{x}}_{k} $ is Gaussian, and contains no information of the past observations $\mathbf{x}_{0,\ldots,k-1}$, i.e., $I \left( \mathbf{x}_k - \overline{\mathbf{x}}_{k}; \mathbf{x}_{0,\ldots,k-1} \right) = 0$. Furthermore, we now present an alternative perspective to look at the term $I \left( \mathbf{x}_k - \overline{\mathbf{x}}_{k}; \mathbf{x}_{0,\ldots,k-1} \right)$ in the following proposition, whose proof is omitted.

\begin{proposition} With $\overline{\mathbf{x}}_{k} = f_{k} \left( \mathbf{x}_{0,\ldots,k-1} \right)$, it always holds that
	\begin{flalign}
	&I \left( \mathbf{x}_k - \overline{\mathbf{x}}_{k}; \mathbf{x}_{0,\ldots,k-1} \right) \nonumber \\
	&~~~~ = I \left( \mathbf{x}_{k} - \overline{\mathbf{x}}_{k} ; \mathbf{x}_{0} - \overline{\mathbf{x}}_{0}, \ldots, \mathbf{x}_{k-1} - \overline{\mathbf{x}}_{k-1} \right).
	\end{flalign}
\end{proposition}


Hence, the condition that $I \left( \mathbf{x}_k - \overline{\mathbf{x}}_{k}; \mathbf{x}_{0,\ldots,k-1} \right) = 0$ is equivalent to that 
\begin{flalign}
I \left( \mathbf{x}_{k} - \overline{\mathbf{x}}_{k} ; \mathbf{x}_{0} - \overline{\mathbf{x}}_{0}, \ldots, \mathbf{x}_{k-1} - \overline{\mathbf{x}}_{k-1} \right) = 0,
\end{flalign}
which means that the current innovation $\mathbf{x}_k - \overline{\mathbf{x}}_{k} $ contains no information of the past innovations. This is a key observation that enables the subsequent analysis in the asymptotic case.

\begin{corollary} \label{MIMOasymp}
	Consider a stochastic process $\left\{ \mathbf{x}_{k} \right\}, \mathbf{x}_{k} \in \mathbb{R}^{n}$. Denote the $1$-step prediction of $\mathbf{x}_{k}$ by $\overline{\mathbf{x}}_{k} = f_{k} \left( \mathbf{x}_{0,\ldots,k-1} \right)$.
	Then, the following prediction error bound always holds:  
	\begin{flalign} \label{MIMOasymp1}
	&\liminf_{k\to \infty} \det \mathrm{E} \left[ \left( \mathbf{x}_{k} - \overline{\mathbf{x}}_{k} \right) \left( \mathbf{x}_{k} - \overline{\mathbf{x}}_{k} \right)^T \right] \nonumber \\
	&~~~~ \geq \liminf_{k\to \infty} \frac{1}{\left( 2 \pi \mathrm{e} \right)^n} 2^{2 h \left( \mathbf{x}_k | \mathbf{x}_{0,\ldots,k-1} \right)},
	\end{flalign}
	where equality holds if and only if $\left\{ \mathbf{x}_{k} - \overline{\mathbf{x}}_{k} \right\}$ is asymptotically white Gaussian.
\end{corollary}

Corollary~\ref{MIMOasymp} follows directly from Theorem~\ref{MIMOprediction} by taking $\liminf_{k\to \infty}$ on both sides of \eqref{MIMOprediction1}. Correspondingly, equality in \eqref{MIMOasymp1} holds if and only if $\mathbf{x}_{k} - \overline{\mathbf{x}}_{k}$ is Gaussian and 
\begin{flalign}
&I \left( \mathbf{x}_k - \overline{\mathbf{x}}_{k}; \mathbf{x}_{0,\ldots,k-1} \right) \nonumber \\
&~~~~ = I \left( \mathbf{x}_{k} - \overline{\mathbf{x}}_{k} ; \mathbf{x}_{0} - \overline{\mathbf{x}}_{0}, \ldots, \mathbf{x}_{k-1} - \overline{\mathbf{x}}_{k-1} \right)= 0,
\end{flalign}
as $k\to \infty$. Since $\mathbf{x}_{k} - \overline{\mathbf{x}}_{k}$ being Gaussian as $k\to \infty$ means that $\mathbf{x}_{k} - \overline{\mathbf{x}}_{k}$ is asymptotically Gaussian, and noting that 
\begin{flalign}
I \left( \mathbf{x}_{k} - \overline{\mathbf{x}}_{k} ; \mathbf{x}_{0} - \overline{\mathbf{x}}_{0}, \ldots, \mathbf{x}_{k-1} - \overline{\mathbf{x}}_{k-1} \right)= 0 \nonumber
\end{flalign}
as $k\to \infty$ is equivalent to that $\mathbf{x}_{k} - \overline{\mathbf{x}}_{k}$ is asymptotically white, equality in \eqref{MIMOasymp1} holds if and only if $\left\{ \mathbf{x}_{k} - \overline{\mathbf{x}}_{k} \right\}$ is asymptotically white Gaussian.

We now know that in order to achieve the prediction error bounds asymptotically, the necessary and sufficient condition is that the innovation is asymptotically white Gaussian, mandating the optimal, minimum-variance predictor to be an innovation Gaussianizing-whitening predictor; refer to \cite{FangITW19arxiv} for further discussions on this topic.

In fact, all the previous prediction error bounds are valid for $1$-step prediction. More generally, for $m$-step prediction ($m$ is any positive integer), the following results can be obtained.

\begin{theorem} \label{mstep}
	Consider a stochastic process $\left\{ \mathbf{x}_{k} \right\}, \mathbf{x}_{k} \in \mathbb{R}^{n}$. Denote the $m$-step prediction of $\mathbf{x}_{k}$ by $\overline{\mathbf{x}}_{k} = f_{k} \left( \mathbf{x}_{0,\ldots,k-m} \right)$.
	Then, the following prediction error bound always holds:  
	\begin{flalign} 
	\det \mathrm{E} \left[ \left( \mathbf{x}_{k} - \overline{\mathbf{x}}_{k} \right) \left( \mathbf{x}_{k} - \overline{\mathbf{x}}_{k} \right)^T \right]
	\geq \frac{1}{\left( 2 \pi \mathrm{e} \right)^n} 2^{2 h \left( \mathbf{x}_k | \mathbf{x}_{0,\ldots,k-m} \right)},
	\end{flalign}
	where equality holds if and only if $\mathbf{x}_{k} - \overline{\mathbf{x}}_{k}$ is Gaussian and $I \left( \mathbf{x}_k - \overline{\mathbf{x}}_{k}; \mathbf{x}_{0,\ldots,k-m} \right) = 0$.
\end{theorem}

To compare with the $l$-step ($1 \leq l \leq m-1$) prediction error bound, note that
\begin{flalign} 
h \left( \mathbf{x}_k | \mathbf{x}_{0,\ldots,k-m} \right) 
&= h \left( \mathbf{x}_k | \mathbf{x}_{0,\ldots,k-\ell} \right) \nonumber \\
&~~~~ + I \left( \mathbf{x}_k ; \mathbf{x}_{k-m+1,\ldots,k-\ell} | \mathbf{x}_{0,\ldots,k-m} \right). \nonumber
\end{flalign}
As such,
\begin{flalign} 
h \left( \mathbf{x}_k | \mathbf{x}_{0,\ldots,k-m} \right) 
&\geq h \left( \mathbf{x}_k | \mathbf{x}_{0,\ldots,k-m+1} \right) \nonumber \\
&\geq \cdots \geq h \left( \mathbf{x}_k | \mathbf{x}_{0,\ldots,k-1} \right).
\end{flalign}
That is to say, the prediction error bound will not decrease as the prediction step increases. On the other hand,
\begin{flalign} 
h \left( \mathbf{x}_k | \mathbf{x}_{0,\ldots,k-m} \right) 
\leq h \left( \mathbf{x}_k \right).
\end{flalign}
Stated alternatively, in the worst case, the error bound is given by the entropy power \cite{Cov:06} of the data point to be predicted, defined as
\begin{flalign} 
\frac{1}{2 \pi \mathrm{e}} 2^{\frac{2}{n} h \left( \mathbf{x}_k \right)}.
\end{flalign}
To sum up, $h \left( \mathbf{x}_k | \mathbf{x}_{0,\ldots,k-m} \right) $ is lower bounded and upper bounded as
\begin{flalign} 
h \left( \mathbf{x}_k | \mathbf{x}_{0,\ldots,k-1} \right)
\leq h \left( \mathbf{x}_k | \mathbf{x}_{0,\ldots,k-m} \right) 
\leq h \left( \mathbf{x}_k \right).
\end{flalign}

When $k \to \infty$, the next corollary follows for $m$-step prediction.

\begin{corollary}
	Consider a stochastic process $\left\{ \mathbf{x}_{k} \right\}, \mathbf{x}_{k} \in \mathbb{R}^{n}$. Denote the $m$-step prediction of $\mathbf{x}_{k}$ by $\overline{\mathbf{x}}_{k} = f_{k} \left( \mathbf{x}_{0,\ldots,k-m} \right)$.
	Then, the following prediction error bound always holds:  
	\begin{flalign} \label{MIMOasymp1m}
	&\liminf_{k\to \infty} \det \mathrm{E} \left[ \left( \mathbf{x}_{k} - \overline{\mathbf{x}}_{k} \right) \left( \mathbf{x}_{k} - \overline{\mathbf{x}}_{k} \right)^T \right] \nonumber \\
	&~~~~ \geq \liminf_{k\to \infty} \frac{1}{\left( 2 \pi \mathrm{e} \right)^n} 2^{2 h \left( \mathbf{x}_k | \mathbf{x}_{0,\ldots,k-m} \right)},
	\end{flalign}
	where equality holds if and only if $\left\{ \mathbf{x}_{k} - \overline{\mathbf{x}}_{k} \right\}$ is asymptotically Gaussian and colored up to the order of $m-1$.
\end{corollary}

Note that we say a stochastic process is (asymptotically) colored up to the order of $m-1$ if $\mathbf{x}_{k}$ is (asymptotically) independent of $\mathbf{x}_{k-m,k-m-1, \ldots}$. As in Corollary~\ref{MIMOasymp}, it can be shown that equality in \eqref{MIMOasymp1m} holds if and only if $\mathbf{x}_{k} - \overline{\mathbf{x}}_{k}$ is asymptotically Gaussian and
\begin{flalign}
I \left( \mathbf{x}_{k} - \overline{\mathbf{x}}_{k} ; \mathbf{x}_{0} - \overline{\mathbf{x}}_{0}, \ldots, \mathbf{x}_{k-m} - \overline{\mathbf{x}}_{k-m} \right) = 0 \nonumber
\end{flalign}
as $k\to \infty$, which is in turn equivalent to that $\mathbf{x}_{k} - \overline{\mathbf{x}}_{k}$ is asymptotically colored up to order $m-1$. Clearly, when $m=1$, being ``colored up to order $\left( m-1 = 0 \right)$" is equivalent to being white, reducing to the case of $1$-step prediction.

\subsection{Connections with the Kolmogorov--Szeg\"o and Wiener--Masani Formulae}

When the process to be predicted is asymptotically stationary and Gaussian, the prediction error bounds can be related to the Kolmogorov--Szeg\"o formula and Wiener--Masani formula. In order to show this, let us first present the next corollary.

\begin{corollary} \label{Gaussian}
	Consider an asymptotically stationary stochastic process $\left\{ \mathbf{x}_{k} \right\}, \mathbf{x}_{k} \in \mathbb{R}^{n}$. Denote the $1$-step prediction of $\mathbf{x}_{k}$ by $\overline{\mathbf{x}}_{k} = f_{k} \left( \mathbf{x}_{0,\ldots,k-1} \right)$.
	Then, 
	\begin{flalign} 
	\liminf_{k\to \infty} \det \mathrm{E} \left[ \left( \mathbf{x}_{k} - \overline{\mathbf{x}}_{k} \right) \left( \mathbf{x}_{k} - \overline{\mathbf{x}}_{k} \right)^T \right] 
	\geq \frac{1}{\left( 2 \pi \mathrm{e} \right)^n} 2^{2 h_{\infty} \left( \mathbf{x} \right)}.
	\end{flalign}
	where $h_{\infty} \left( \mathbf{x} \right)$ denotes the entropy rate \cite{Cov:06} of $\left\{ \mathbf{x}_{k} \right\}$, and equality holds if and only if $\left\{ \mathbf{x}_{k} - \overline{\mathbf{x}}_{k} \right\}$ is asymptotically white Gaussian.
\end{corollary}

Corollary~\ref{Gaussian} follows directly from Corollary~\ref{MIMOasymp} by noting that for an asymptotically stationary process $\left\{ \mathbf{x}_{k} \right\}$, we have \cite{Cov:06} 
\begin{flalign} 
\liminf_{k\to \infty} h \left( \mathbf{x}_k | \mathbf{x}_{0,\ldots,k-1} \right) 
= \lim_{k\to \infty} h \left( \mathbf{x}_k | \mathbf{x}_{0,\ldots,k-1} \right)
= h_{\infty} \left( \mathbf{x} \right). \nonumber
\end{flalign}

Moreover, suppose that $\left\{ \mathbf{x}_{k} \right\}$ is asymptotically stationary Gaussian with asymptotic power spectrum $ \Phi_{\mathbf{x}} \left( \omega \right)$. Then, according to \cite{Cov:06}, we have
\begin{flalign} 
h_{\infty} \left( \mathbf{x} \right)
= \frac{1}{2 \mathrm{\pi}} \int_{-\mathrm{\pi}}^{\mathrm{\pi}} \log \sqrt{\left(2 \pi \mathrm{e}\right)^n \det \Phi_{ \mathbf{x} } \left( \omega \right) } \mathrm{d}\omega.  \nonumber
\end{flalign}
Hence, the following prediction error bound holds:  
\begin{flalign} \label{WienerMasani}
&\liminf_{k\to \infty} \det \mathrm{E} \left[ \left( \mathbf{x}_{k} - \overline{\mathbf{x}}_{k} \right) \left( \mathbf{x}_{k} - \overline{\mathbf{x}}_{k} \right)^T \right] \nonumber \\
&~~~~ \geq 2^{\frac{1}{2 \mathrm{\pi}}\int_{-\mathrm{\pi}}^{\mathrm{\pi}}{\log \det \Phi_{ \mathbf{x} } \left( \omega \right)  \mathrm{d}\omega }},
\end{flalign}
in which equality holds if and only if $\left\{ \mathbf{x}_{k} - \overline{\mathbf{x}}_{k} \right\}$ is asymptotically white Gaussian.

This coincides with the Wiener--Masani formula. Moreover, as in the Wiener--Masani formula, equality in \eqref{WienerMasani} is achieved with a linear whitening predictor. To see this, note that since $\overline{\mathbf{x}}_{k}$ is Gaussian, $\mathbf{x}_{k} - \overline{\mathbf{x}}_{k}$ will be Gaussian with a linear filter. Hence, with a linear whitening predictor, $\left\{ \mathbf{x}_{k} - \overline{\mathbf{x}}_{k} \right\}$ is asymptotically white Gaussian; in this case, note also that
\begin{flalign} \nonumber
&\liminf_{k\to \infty} \det \mathrm{E} \left[ \left( \mathbf{x}_{k} - \overline{\mathbf{x}}_{k} \right) \left( \mathbf{x}_{k} - \overline{\mathbf{x}}_{k} \right)^T \right] \nonumber \\
&~~~~ = \lim_{k\to \infty} \det \mathrm{E} \left[ \left( \mathbf{x}_{k} - \overline{\mathbf{x}}_{k} \right) \left( \mathbf{x}_{k} - \overline{\mathbf{x}}_{k} \right)^T \right], \nonumber
\end{flalign}
and hence
\begin{flalign}
&\liminf_{k\to \infty} \det \mathrm{E} \left[ \left( \mathbf{x}_{k} - \overline{\mathbf{x}}_{k} \right) \left( \mathbf{x}_{k} - \overline{\mathbf{x}}_{k} \right)^T \right] \nonumber \\
&~~~~ = 2^{\frac{1}{2 \mathrm{\pi}}\int_{-\mathrm{\pi}}^{\mathrm{\pi}}{\log \det \Phi_{ \mathbf{x} } \left( \omega \right)  \mathrm{d}\omega }}.
\end{flalign}

On the other hand, when $n=1$, \eqref{WienerMasani} reduces to
\begin{flalign}
\liminf_{k\to \infty} \mathrm{E} \left[ \left( \mathbf{x}_{k} - \overline{\mathbf{x}}_{k} \right)^2 \right] 
\geq 2^{\frac{1}{2 \mathrm{\pi}}\int_{-\mathrm{\pi}}^{\mathrm{\pi}}{\log S_{ \mathbf{x} } \left( \omega \right)  \mathrm{d}\omega }},
\end{flalign}
which coincides with the Kolmogorov--Szeg\"o formula.

The Wiener--Masani and Kolmogorov--Szeg\"o formulae are valid for $1$-step prediction. We now obtain the generalized formulae for $m$-step prediction.

\begin{corollary}
	Consider an asymptotically stationary stochastic process $\left\{ \mathbf{x}_{k} \right\}, \mathbf{x}_{k} \in \mathbb{R}^{n}$. Denote the $m$-step prediction of $\mathbf{x}_{k}$ by $\overline{\mathbf{x}}_{k} = f_{k} \left( \mathbf{x}_{0,\ldots,k-m} \right)$.
	Then,  
	\begin{flalign} \label{GaussianM}
	&\liminf_{k\to \infty} \det \mathrm{E} \left[ \left( \mathbf{x}_{k} - \overline{\mathbf{x}}_{k} \right) \left( \mathbf{x}_{k} - \overline{\mathbf{x}}_{k} \right)^T \right] \nonumber \\
	&~~~~ \geq 2^{\frac{1}{2 \mathrm{\pi}}\int_{-\mathrm{\pi}}^{\mathrm{\pi}}{\log \det \Phi_{ \mathbf{x} } \left( \omega \right)  \mathrm{d}\omega }} \nonumber \\
	&~~~~~~~~  + \lim_{k \to \infty} \frac{1}{\left( 2 \pi \mathrm{e} \right)^n} 2^{2  I \left( \mathbf{x}_k ; \mathbf{x}_{k-m+1,\ldots,k-1} | \mathbf{x}_{0,\ldots,k-m} \right)}. 
	\end{flalign}
	Herein, equality holds if and only if $\left\{ \mathbf{x}_{k} - \overline{\mathbf{x}}_{k} \right\}$ is asymptotically Gaussian and colored up to the order of $m-1$.
\end{corollary}

	
Clearly, there will be an additional term of (conditional) mutual information in the case of $m$-step prediction compared with $1$-step prediction; see discussions after Theorem~\ref{mstep} for properties of this additional term. 
	
On the other hand, the corresponding formulae for $1$-step prediction of non-Gaussian processes can also be obtained, as shown in the following corollary (its proof is omitted). 

\begin{corollary}
	Consider an asymptotically stationary stochastic process $\left\{ \mathbf{x}_{k} \right\}, \mathbf{x}_{k} \in \mathbb{R}^{n}$ with asymptotic power spectrum $ \Phi_{\mathbf{x}} \left( \omega \right)$. Denote the $1$-step prediction of $\mathbf{x}_{k}$ by $\overline{\mathbf{x}}_{k} = f_{k} \left( \mathbf{x}_{0,\ldots,k-1} \right)$.
	Then,  	
	\begin{flalign} 
	&\liminf_{k\to \infty} \det \mathrm{E} \left[ \left( \mathbf{x}_{k} - \overline{\mathbf{x}}_{k} \right) \left( \mathbf{x}_{k} - \overline{\mathbf{x}}_{k} \right)^T \right] \nonumber \\
	&~~~~ \geq \left[ 2^{- 2 J_{\infty} \left( \mathbf{x} \right)} \right] 2^{\frac{1}{2 \mathrm{\pi}}\int_{-\mathrm{\pi}}^{\mathrm{\pi}}{\log \det \Phi_{ \mathbf{x} } \left( \omega \right)  \mathrm{d}\omega }},
	\end{flalign}
	where $J_{\infty} \left( \mathbf{x} \right)$ denotes the negentropy rate \cite{fang2017towards} of $\left\{ \mathbf{x}_{k} \right\}$, $J_{\infty} \left( \mathbf{x} \right) \geq 0$, and $J_{\infty} \left( \mathbf{x} \right) = 0$ if and only if $\left\{ \mathbf{x}_{k} \right\}$ is Gaussian.
    Herein, equality holds if and only if $\left\{ \mathbf{x}_{k} - \overline{\mathbf{x}}_{k} \right\}$ is asymptotically white Gaussian.
\end{corollary}



Herein, negentropy rate is a measure of non-Gaussianity for asymptotically stationary processes, which grows larger as the process to be predicted becomes less Gaussian; see \cite{fang2017towards} for more details of its properties. Accordingly, the bounds will decrease as the process to be predicted becomes less Gaussian.

\subsection{Further Discussions} \label{discussion}

\subsubsection{Relation to data fitting}

Note that the subscript $k$ in $\mathbf{x}_{k}$ does not necessarily denote ``time", but may more generally denote the indices of the data points. As such, the time series estimation and prediction problem might more generally be viewed as a data fitting and extrapolation problem. Accordingly, the generic variance bounds on estimation and prediction errors may then correspond to generic variance bounds on fitting and extrapolation errors.

\subsubsection{Discrete-time counterpart to Duncan's formula}

In a way, our bounds obtained in Section~\ref{estimation} can be viewed as the discrete-time counterparts to Duncan's formula \cite{duncan1970calculation}, which provides lower bounds on the estimation errors for continuous-time processes. This said, we are still investigating the potential deeper and more mathematical connections with Duncan's formula and the many interesting and insightful works that ensued (see, e.g., \cite{guo2005mutual, weissman2013directed, guo2013interplay, weissmannee378, tanaka2017optimal, dytso2017view} and the references therein; see also \cite{FangCDC17, FangACC18, FangCDC18} for some relevant results that might be useful in establishing such connections.).

\section{Conclusions}

	In this paper, we have derived generic bounds on the variances of estimation and prediction errors in time series analysis using information theory, and the bounds are applicable to any causal estimators/predictors while the processes to be estimated/predicted can have arbitrary distributions. 
	Moreover, it is discovered that the necessary and sufficient condition to achieve the fundamental prediction error bounds asymptotically is that the innovation is asymptotically white Gaussian.
	When it comes to Gaussian processes and $1$-step prediction, our bounds reduce to the Kolmogorov--Szeg\"o formula and Wiener--Masani formula in linear prediction theory.



\bibliographystyle{IEEEtran}
\bibliography{references}

\end{document}